\documentclass[12pt]{article}
\usepackage{mathrsfs}
\usepackage[T1]{fontenc}
\usepackage{mathpazo}
\usepackage{setspace}
\usepackage{amsfonts}
\usepackage{amssymb}
\usepackage{epsfig}
\usepackage{latexsym}
\usepackage{color}
\usepackage{graphicx}
\usepackage{nicefrac}
\usepackage[latin1]{inputenc}
\usepackage{pstricks}
\usepackage{slashed}
\usepackage{multirow}
\def\hybrid{\topmargin -30pt    \oddsidemargin 0pt %%%%%%%%%%%%%% Archive-30pt
        \headheight 0pt \headsep 0pt
        \textwidth 6.25in       % A4 paper
        \textheight 9.5in       % A4 paper
        \marginparwidth .875in
        \parskip 5pt plus 1pt   \jot = 1.5ex}

\hybrid

\def\baselinestretch{1.2}

\catcode`\@=11

\def\marginnote#1{}
\newcount\hour
\newcount\minute
\newtoks\amorpm
\hour=\time\divide\hour by60
\minute=\time{\multiply\hour by60 \global\advance\minute by-\hour}
\edef\standardtime{{\ifnum\hour<12 \global\amorpm={am}%
        \else\global\amorpm={pm}\advance\hour by-12 \fi
        \ifnum\hour=0 \hour=12 \fi
        \number\hour:\ifnum\minute<10 0\fi\number\minute\the\amorpm}}
\edef\militarytime{\number\hour:\ifnum\minute<10 0\fi\number\minute}
\def\draftlabel#1{{\@bsphack\if@filesw {\let\thepage\relax
   \xdef\@gtempa{\write\@auxout{\string
      \newlabel{#1}{{\@currentlabel}{\thepage}}}}}\@gtempa
   \if@nobreak \ifvmode\nobreak\fi\fi\fi\@esphack}
        \gdef\@eqnlabel{#1}}
\def\@eqnlabel{}
\def\@vacuum{}
\def\draftmarginnote#1{\marginpar{\raggedright\scriptsize\tt#1}}

\def\draft{\oddsidemargin -.5truein
        \def\@oddfoot{\sl preliminary draft \hfil
        \rm\thepage\hfil\sl\today\quad\militarytime}
        \let\@evenfoot\@oddfoot \overfullrule 3pt
        \let\label=\draftlabel
        \let\marginnote=\draftmarginnote
   \def\@eqnnum{(\theequation)\rlap{\kern\marginparsep\tt\@eqnlabel}%
\global\let\@eqnlabel\@vacuum}  }

\def\draft2{
        \def\@oddfoot{\sl preliminary draft \hfil
        \rm\thepage\hfil\sl\today\quad\militarytime}
        \let\@evenfoot\@oddfoot \overfullrule 3pt
        \let\label=\draftlabel
        \let\marginnote=\draftmarginnote
   \def\@eqnnum{(\theequation)\rlap{\kern\marginparsep\tt\@eqnlabel}%
\global\let\@eqnlabel\@vacuum}  }

\def\preprint{\twocolumn\sloppy\flushbottom\parindent 2em
        \leftmargini 2em\leftmarginv .5em\leftmarginvi .5em
        \oddsidemargin -.5in    \evensidemargin -.5in
        \columnsep .4in \footheight 0pt
        \textwidth 10.in        \topmargin  -.4in
        \headheight 12pt \topskip .4in
        \textheight 6.9in \footskip 0pt
        \def\@oddhead{\thepage\hfil\addtocounter{page}{1}\thepage}
        \let\@evenhead\@oddhead \def\@oddfoot{} \def\@evenfoot{} }

\def\numberbysection{\@addtoreset{equation}{section}
        \def\theequation{\thesection.\arabic{equation}}}

\def\underline#1{\relax\ifmmode\@@underline#1\else
        $\@@underline{\hbox{#1}}$\relax\fi}

\def\titlepage{\@restonecolfalse\if@twocolumn\@restonecoltrue\onecolumn
     \else \newpage \fi \thispagestyle{empty}\c@page\z@
        \def\thefootnote{\fnsymbol{footnote}} }

\def\endtitlepage{\if@restonecol\twocolumn \else \newpage \fi
        \def\thefootnote{\arabic{footnote}}
        \setcounter{footnote}{0}}  %\c@footnote\z@ }

\catcode`@=12
\relax

\def\figcap{\section*{Figure Captions\markboth
        {FIGURECAPTIONS}{FIGURECAPTIONS}}\list
        {Figure \arabic{enumi}:\hfill}{\settowidth\labelwidth{Figure
999:}
        \leftmargin\labelwidth
        \advance\leftmargin\labelsep\usecounter{enumi}}}
 \relax
\def\tablecap{\section*{Table Captions\markboth
        {TABLECAPTIONS}{TABLECAPTIONS}}\list
        {Table \arabic{enumi}:\hfill}{\settowidth\labelwidth{Table
999:}
        \leftmargin\labelwidth
        \advance\leftmargin\labelsep\usecounter{enumi}}}
 \relax
\def\reflist{\section*{References\markboth
        {REFLIST}{REFLIST}}\list
        {[\arabic{enumi}]\hfill}{\settowidth\labelwidth{[999]}
        \leftmargin\labelwidth
        \advance\leftmargin\labelsep\usecounter{enumi}}}
 \relax
\makeatletter
\newcounter{pubctr}
\def\publist{\@ifnextchar[{\@publist}{\@@publist}}
\def\@publist[#1]{\list
        {[\arabic{pubctr}]\hfill}{\settowidth\labelwidth{[999]}
        \leftmargin\labelwidth
        \advance\leftmargin\labelsep
        \@nmbrlisttrue\def\@listctr{pubctr}
        \setcounter{pubctr}{#1}\addtocounter{pubctr}{-1}}}
\def\@@publist{\list
        {[\arabic{pubctr}]\hfill}{\settowidth\labelwidth{[999]}
        \leftmargin\labelwidth
        \advance\leftmargin\labelsep
        \@nmbrlisttrue\def\@listctr{pubctr}}}
 \relax
\makeatother

\def\be{\begin{equation}}
\def\ee{\end{equation}}
\def\ba{\begin{eqnarray}}
\def\ea{\end{eqnarray}}

\def\del{\partial}

\def\r{\rho}
\def\a{\alpha}

\def\b{\beta}

\def\G{\Gamma}
\def\d{\delta}

\def\e{\epsilon}

\def\th{\theta}

\def\m{\mu}
\def\n{\nu}
\def\om{\omega}
\def\Om{\Omega}
\def\l{\lambda}
\def\L{\Lambda}
\def\s{\sigma}
\def\S{\Sigma}

\def\cN{{\cal N}}

\def\no{\noindent}

\def\qq{\qquad}

\def\IR{\relax{\rm I\kern-.18em R}}

\def\inv{^{\raise.0ex\hbox{${\scriptscriptstyle -}$}\kern-.05em 1}}

\def \ha {{\frac{1}{2}}}

\def \ov {\over}

\def\diag{{\rm diag}}

\begin{document}
%\draft2

%\renewcommand{\theequation}{\arabic{equation}}
%\renewcommand{\theequation}{\thesection.\arabic{equation}}

\renewcommand{\theequation}{\thesection.\arabic{equation}}
\csname @addtoreset\endcsname{equation}{section}

\begin{titlepage}
\begin{center}

%{}\hfill QMUL-PH-10-

\phantom{xx}
\vskip 0.5in

{\large \bf On non-abelian T-dual geometries with Ramond  fluxes}

\vskip 0.45in

{\bf Konstadinos Sfetsos}${}^{1}$ \phantom{x}and\phantom{x}{\bf Daniel C. Thompson}${}^{2}$
\vskip 0.1in

${}^1$Department of Engineering Sciences, University of Patras,\\
26110 Patras, Greece\\
{\tt sfetsos@upatras.gr}

\vskip .2in

${}^2$  Theoretische Natuurkunde, Vrije Universiteit Brussel, and \\
International Solvay Institutes\\
Pleinlann 2, B-1050, Brussels, Belgium \\
{\tt dthompson@tena4.vub.ac.be}

\end{center}

\vskip .4in

\centerline{\bf Abstract}

\no
We show how to implement T-duality along non-abelian isometries
in backgrounds with non-vanishing Ramond fields.  When the 
dimension of the isometry group is odd (even) the duality swaps 
(preserves) the chirality of the theory.  In certain cases a 
non-abelian duality can result in a massive type-IIA background.  
We provide two examples by dualising 
   $SU(2)$ isometry subgroups in  $AdS_5 \times S^5$ 
and $AdS_3\times S^3 \times T^4$. 
The resultant dual geometries inherit the original $AdS$ factors
but have transverse spaces with reduced isometry and
 preserve only half of the original supersymmetry. 
 The non-abelian dual of $AdS_5 \times S^5$ has an M-theory lift which is related to
 the gravity duals of $\cN=2$ superconformal
theories. We comment on a possible interpretation of this as a high spin limit.

\end{titlepage}
\vfill
\eject

%\def\baselinestretch{1.2}
%\baselineskip 10 pt
%\noindent

\tableofcontents

\def\baselinestretch{1.2}
\baselineskip 20 pt
\no

\newcommand{\eqn}[1]{(\ref{#1})}

\section{Introduction}

T-duality is one of the most striking features of string theory and,
though its discovery dates back a quarter of a century, it  continues to
provide us new insight into the
nature of string theory.

\no
An interesting recent development has been the use of T-duality in
conjunction with the AdS/CFT correspondence to explain the
amplitude/Wilson loop  connection of ${\cal N } =4$ gauge theory   \cite{Berkovits:2008ic,Alday:2007hr}. This application
required one to perform both traditional bosonic and novel fermionic
T-dualities in an $AdS_5\times S^5$ background supported by Ramond flux.
These fluxes are of immense importance since they support the gravitational
backgrounds corresponding to the various Dp-branes that have been instrumental to
all fundamental theoretical developments ranging from U-duality to
the AdS/CFT correspondence.
Given the importance of such Ramond backgrounds it is natural to ask what other
   applications could T-duality have in this context.

   \no
In this paper we shall discuss the extension of T-duality to the case of isometries which form
   a non-abelian group structure \cite{delaossa:1992vc}.
In comparison
   to the abelian T-duality this non-abelian generalisation has remained
   rather poorly understood in spite of a substantial body of work, e.g.
 \cite{Giveon:1993ai}-\cite{Sfetsos:1994vz}.
    One reason for this is that global topological issues on the world sheet make
    it hard to establish the duality as an exact symmetry of a
Conformal Field Theory (CFT) \cite{Giveon:1993ai}.
 Nonetheless, one may view these transformations as a map between related CFT's or as
 generating new effective backgrounds. In that respect it has been recently shown that
non-abelian T-duality can be considered as an effective theory describing consistent sectors
of infinite highest weight representations, i.e. infinitely large spins, of certain
parent theories involving Neveu--Schwarz (NS) fields only \cite{Polychronakos:2010nk}.

\no
In this paper we move a step forward in understanding non-abelian T-duality structures in
string theory by examining how Ramond--Ramond (RR) field strengths transform.
Since this presents technical as well as conceptual difficulties we restrict ourselves to
the case where the directly relevant part of the Neveu--Schwarz sector in the supergravity
background takes the form of a Principal Chiral Model.

\no
 For abelian T-duality the extension of the Buscher rules \cite{Buscher:1987sk}
  to such Ramond backgrounds was first achieved from a target space perspective \cite{Bergshoeff:1995as}
  (see also  \cite{Hassan:1999bv,Hassan:1999mm}), and then established from a world
   sheet perspective in the Green--Schwarz formalism \cite{Cvetic:1999zs,Kulik:2000nr}
   and more recently in the pure spinor formalism \cite{Benichou:2008it} and as a canonical
   transformation \cite{Sfetsos:2010pl}.

\no
 The key to understanding the behaviour of RR-fields is the observation that left and
 right movers transform differently under T-duality, i.e.
  $\partial_\pm \tilde{X}^I = (A_\pm )^I{}_J\partial_\pm  X^J$,
for certain matrices $A_\pm $ that depend on the metric and
NS antisymmetric tensor \cite{Hassan:1999bv,Hassan:1999mm}.
In particular, the left and right movers of the dual theory couple to different target space vielbeins  related to each other by a local Lorentz transformation given
by $\Lambda = A_+^{-1}A_-$.   To express the dual theory in a single frame one uses this transformation to rotate the left movers back to the same frame as the right movers.  Doing so induces some transformation $\Omega$ on the target space spinor indices. 
Since the RR-field strengths, together with the dilaton,
 can be combined into a bispinor $P_{\a \b}$, they undergo a transformation of the
 form $\hat{P} = P \Omega\inv$   from which the transformation of the individual
RR-fluxes can be obtained.

\no
We will see that the above argument essentially holds for the case of non-abelian
T-duality however with some important modifications. Firstly, the induced local Lorentz
transformation is somewhat more complicated since it may depend on the "internal" directions,
i.e. those being dualised.  Furthermore, we will see that the determinant of the Lorentz
 transformation $\Lambda$ may be either plus or minus one, depending on the dimensionality of the
 isometry group being dualised.  Starting with a type-IIB string background the resultant
 theory will remain of type-IIB for an even-dimensional isometry group, but for an
odd-dimensional isometry group the T-dual is a background of type-IIA supergravity
and vice versa.
 When the starting background is of type-IIB and has an RR three-form with support
in the directions being dualised corresponding to an $SU(2)$ isometry group,
it is possible to end up with a background
of massive type-IIA supergravity, a theory constructed in \cite{Romans:1985tz}.
More generally this may happen when the T-duality transformation involves
$p$-forms and $p$-dimensional groups.

\no
We will apply these ideas to two familiar geometries;
the $AdS_3 \times S^3 \times T^4$ background arising in the near
 horizon limit of the D1-D5 system and the $AdS_5 \times S^5$
background corresponding to the D3-brane near horizon geometry.
In both cases we shall consider the dualisation along isometries forming
an $su(2)$ algebra which generate the left (or right) action on a three-sphere in the geometry.
   The result will be that we find dual geometries in which the $AdS$
factor remains unaltered but the transverse spaces are quite different and have reduced isometry.
   Correspondingly we find that the dual geometries have reduced supersymmetry with,
in both cases, exactly one half of the original supersymmetries preserved.
We show that in the dual geometries the preserved Killing spinors obey
position depended projection conditions. 
We suggest that the fraction of preserved supersymmetry can be established by
considering the action of the spinor-Lorentz-Lie-(Kosmann)  derivative
\cite{Kosmann,oFarrill:1999va, Ortin:2002qb} along the
direction of the isometries about which we dualise. More specifically,
we suggest that the number of Killing spinors of the original geometry for
which this derivative vanishes correspond
to the number of supersymmetries preserved under dualisation.

\no
For the case of the non-abelian T-dual $AdS_5 \times S^5$
background we show that the dual geometry has an M-theory lift which is related to the 
Gaiotto--Maldacena (GM) gravity duals of $\cN=2$ superconformal theories. 
More precisely we find a background which can be obtained as a leading order approximation to a  generic GM background close to the origin.   We provide a possible interpretation of this in terms of a high spin limit of a parent $\cN=2$ gauge theory. 
 
 \no
 The structure of this paper is as follows: in the section 2
 we review the action of non-abelian T-duality in Principal Chiral Models, we establish
 the Lorentz transformation that relates left and right moving frames from which consequently
one may  derive the tranformation rules for the RR fluxes.  In section 3
we consider the dualisation of the D1-D5 near horizon geometry and in section 4 we
examine the $AdS_5\times S^5$ background corresponding the D3-brane near horizon geometry.
We close the paper in section 5 with a short discussion and conclusion.   In the appendix we provide some details of the massive type-IIA supergravity which are relevant to our discussion.

\section{Non-abelian T-duality in Principal Chiral Models }

In this paper we restrict to a particular class of non-abelian T-duality transformations in which
the isometry appears through a
Principal Chiral Model (PCM) \cite{Polyakov:1975rr}-\cite{Luscher:1977rq}.
Whilst the PCMs may not constitute a  solution to string/supergravity theory
on its own, they can often be naturally embedded as integrable
parts, as we shall see, of true supergravity solutions involving D-brane configurations.

\no
A PCM is a class of two-dimensional
$\s$-model with $G_L\times G_R$ global symmetry for a group $G$.
The $\s$-model action  for a group element $g\in G$ is given
by
\be
S(g)= -\int d^{2}\s \, {\rm Tr}(g^{-1}\del_- g g^{-1} \del_+ g)\
\label{pcmac}
\ee
and obviously, it is invariant under a global $G_L\times G_R$ symmetry.
Actually, there ia a much larger classical affine symmetry \cite{Luscher:1977rq, Dolan:1981fq, symaffPCM}
which however will play no r\^ole in our considerations.

\subsection{The T-dual}

We would like to find the non-abelian dual of this action
corresponding to a subgroup $G_L$.\footnote{ For didactic purposes we ignore spectator fields
 (i.e. directions supplementary to those being dualised) coupled to the above PCM action, though they may
easily be included as we do later. }
To achieve this one first gauges some isometry by introducing gauge fields and covariant
derivatives into the $\s$-model. This gauged theory is supplemented with Lagrange
multipliers which enforce that the gauge fields are (locally) pure gauge and means that the
 original $\s$-model is recovered upon gauge fixing them to zero.
If, instead of the Lagrange
multipliers, the  gauge  fields are integrated out one arrives at the T-dual $\s$-model with the Lagrange
multiplier playing the r\^ole of T-dual coordinates. In general, for non-abelian isometries this process
can be complicated and the gauge fixing needs to be done on a case by case basis. However,
for the case at hand, i.e. for PCMs, one may give general expressions.

\no
We choose to gauge the entire $G_L$ isometry with the corresponding action
\be
 S_{\rm nonab}(g,v,A_\pm) =
- \int d^{2} \s \,  {\rm Tr}(g^{-1}D_- g g^{-1} D_+ g) + i {\rm Tr}(v F_{+-})\ ,
\label{gh11}
\ee
where $v$ is the Lagrange multiplier matrix.
The covariant derivatives and the field strength for the gauged fields are given by
\be
D_\pm g  = \del_\pm g -A_\pm g \ ,\qq F_{+-}= \del_+ A_- - \del_-A_+ - [A_+,A_-]\ .
\label{coovder}
\ee
The action above is invariant under the local "left" symmetry
\be
g\to \l ^{-1} g\ ,\quad v\to \l^{-1} v \l\ ,\quad
\quad A_\pm\to \l^{-1}A_\pm \l - \l ^{-1}\del_\pm \l \ ,\quad \l(\s^+,\s^-)\in G\ ,
\ee
as well as the global "right" symmetry $g\to g \L_R$,
where $\L_R\in G$.

\no
The gauge fields in \eqn{gh11} are non-dynamical and  can be eliminated via their equations
of motion. We also gauge fix $\dim (G)$ of the parameters.
As a gauge fixing we choose the group element $g$ to be the identity which is possible since the
group action has no isotropy.
This gauge choice completely gets rid of the parameters
in $G$ and we are left with a $\s$-model solely
for the Lagrange multipliers $v$. The result is a dual action given by
\be
S = \int  d^{2} \s \ \del_+ v_i (M^{-1})^{ij} \del_- v_j \ ,\qq M_{ij} = \d_{ij} + f_{ij}\ ,\quad
f_{ij}\equiv f_{ij}{}^k v_k\ .
\label{vKv}
\ee
From this we may read off the target space metric and NS two-form as the symmetric
and anti-symmetric parts of  $(M^{-1})^{ij}$, respectively. The above action still processes
a global
$G$-symmetry associated to the original right action of the group that is left intact.
The process of integrating out the gauge fields introduces an extra
factor in the path integral measure leading to a non-trivial dilaton in the T-dual theory
given by
\be
e^{-2\Phi}= \det (M)\ .
\ee
Of course this is the true dilaton field only when the PCM is embedded
in a full string background. This is the case in the explicit
examples we consider below.

\subsection{Transforming the RR-fluxes}

In order to see how a corresponding transformation for RR-flux fields
is induced, we first establish how T-duality acts on world-sheet derivatives.
This is properly done in the formulation of non-abelian T-duality
as a canonical transformation in phase space \cite{Curtright:1994be,Lozano:1996sc}.
Introducing local coordinates $\th^a$ on the group manifold $G$
and defining the
left-invariant one-forms as $L =  g^{-1} d  g=  L_a^i d\th^{a} T_{i}$,
we may express the transformation of world-sheet derivatives as \cite{Lozano:1996sc,Sfetsos:2010pl}
\be
\partial_+ v_l = M_{kl} L^k_b \partial_+ \th^b  \ ,\qq
\partial_- v_l = - M_{lk} L^k_b \partial_- \th^b \ .
\ee
The difference in transformation between left and right movers is given by the matrix
\be
\Lambda_i{}^j = - \left(M M^{-1 T} \right)_i{}^j \, .
\label{Lambda}
\ee
Since $M$ and $M^{T}$ commute it can be seen that $\Lambda^{T}= \Lambda^{-1}$
and $|\det \Lambda | =1$. Hence $\Lambda$ defines a local Lorentz transformation
however, not necessarily one connected to the identity (with $\det \Lambda =1 $).
If the dimensionality $\dim(G)$ of $G$ is odd, the Lorentz transformation \eqn{Lambda}
relating left and right movers includes the action of parity whereas
if it is even it does not.
This Lorentz transformation also induces an action
on spinors given by a matrix $\Omega$ obtained by requiring that
\be
\Omega^{-1}  \Gamma^i  \Omega =  \Lambda^i{}_j \Gamma^j\ .
\label{spino1}
\ee
Hence, in the context of type-II superstrings, this transformation will map between
the type-IIA and type-IIB theories when $\dim(G)$ is odd and will preserve
the chirality of the theory when $\dim(G)$ is even.
One may readily check that this corresponds to our received wisdom in the abelianised limit
by taking the structure constants to vanish.  Let us illustrate this in a slightly generalised context
by including some extra un-dualised spectator directions but restricting to flat backgrounds
with spectator fields that have no legs or dependance on the directions being dualised.
The resultant Lorentz transformation
corresponds to performing a T-duality
for an abelian group isomorphic to $U(1)^d$ and is simply given by
\be
\Lambda =  \diag (-\mathbb{1}_{d}, \mathbb{1}_{10-d} ) \ .
\ee
The spinorial representation of this transformation is found by solving \eqn{spino1} and reads
\be
\Omega =  \prod_{i=1}^{d} ( \G^{i} \G_{11})\ ,
\ee
where $\G_{11}=\G^0\G^1 \cdots \G^9$, obeying $\G_{11}\cdot \Gamma_{11}=\mathbb{1}$.
Due to the abelian nature of the group in this case, we may interpret this transformation
as that coming from performing a succession of
standard abelian T-dualities in each different direction.

\no
It is now clear how we may include RR-fields into the discussion.
In the type-IIB case we have odd RR-forms which can be combined into
a bi-spinor as
\be
P = {e^{\Phi}\ov 2} \sum_{n=0}^4{1\ov (2n+1)!}\ \slashed{F}_{2n+1} \ .
\ee
In the type-IIA case we have the similar expression
\be
\hat{P} ={ e^{\Phi}\ov 2} \sum_{n=0}^5 {1\ov (2n)!}\ \slashed{F}_{2n} \ ,
\ee
where we have included a zero form potential corresponding to the mass parameter of massive type-IIA
supergravity.
In addition we have made use of the standard notation
$
{\slashed F}_p =  \G^{\m_1\cdots \m_p} F_{\m_1\cdots \m_p}
$.
% and similarly for $\slashed H$ and $\slashed \del \Phi$.
In the definitions of $P$ and $\hat P$ we have  used the democratic formulation
of type-II supergravities \cite{Bergshoeff:2001pv}.
In this formulation and for Minkowski signature spacetimes
the conditions\footnote{Our conventions for the Hodge dual on a
$p$-form in a $D$-dimensional spacetime are that
\be
(\star F_p)_{\m_{p+1}\cdots \m_D}  = {1\ov p!} \sqrt{|g|}\ \e_{\m_1\cdots \m_D} F_p^{\m_1\cdots \m_p}\ ,
\ee
where $\e_{01\dots 9} =1$.  With this we have the useful identity $\star\star F_p = s (-1)^{p(D-p)}
F_p$, where $s$ is the signature of spacetime which we take to be mostly plus.
}
\be
 F_{2n} = (-1)^n\star F_{10-2n}\ ,
\label{hdhi1}
\ee
for type-IIA
and
\be
F_{2n+1} = (-1)^n\star F_{9-2n}\ ,
\label{hdhi2}
\ee
for type-IIB should be imposed so that one remains with the right degrees of freedom.
However, when we check our solutions to supergravity we shall, in general,
work with the standard formulations of type-II supergravities in which
no higher forms than five appear.

\no
The above fluxes transform according to
\be
\hat P = P \Om\inv\ ,
\label{ppom}
\ee
which is a rule of the same form as in the abelian case.
That this is right transformation rule can be seen from either the
space time arguments of \cite{Hassan:1999bv,Hassan:1999mm}
or the world sheet argument using the pure spinor superstring of
\cite{Benichou:2008it,Sfetsos:2010pl}.
The details of the matrix $\Omega$ corresponding to cases of non-abelian T-duality are
considerably more complicated than the abelian case. Some of the structure of the matrix  $\Om$ is easy to infer.
When the group with respect to which we T-dualize is an odd-dimensional one, $\Om$ starts with $\G_{11}$
to be followed by a linear combination of products of an odd number of Gamma-matrices. If, instead,
the dimensionality of the group is even, then $\G_{11}$ is omitted and the linear combination is performed
with products of an even number of Gamma-matrices.\footnote{In practice, because the fermions
of the theory are Weyl the $\Omega$ that we use
to perform the transformation of the fluxes is  actually in the Weyl representation
and $\G_{11}$ contributes only an overall sign which can be taken care of by a judicial choice of conventions. }

\no
Finally, note that assuming that we start with a type-IIA supergravity background in which \eqn{hdhi1}
is satisfied, then after T-duality the conditions \eqn{hdhi2} if we end up with a
type-IIB background or again
\eqn{hdhi1} if we remain within type-IIA, should be satisfied automatically. A similar
statement holds if we start from a type-IIB background.

\section{Non-abelian T-dual in the D1-D5 near horizon}

As a first example we consider the $AdS_3\times S^3\times T^4$ geometry that arises as the near horizon limit of the D1-D5 brane system.  The type-IIB supergravity background consist of a metric
\be
ds^2 = ds^2({\rm AdS_3}) + ds^2({\rm S^3}) + ds^2({\rm T^4})\ ,
\ee
where the normalization is such $R_{\m\n} = \mp \ha g_{\m\n}$ for the $AdS_3$ and $S^3$
factors, respectively, supported by the Ramond flux
\be
F_3 = {\rm Vol}(AdS_3) + {\rm Vol}(S^3)\ ,
\label{adsfl}
\ee
whereas the dilaton $\Phi =0$.\footnote{Due to S-duality there exists a one parameter family of solutions with both RR and NS three-form fluxes turned on. However for this paper we only wish to consider the case with just RR flux.}
Note that we have completely absorbed all constant factors by appropriate rescalings.
The presence of $S^3$ indicates a global $SO(4) \simeq
SU(2)_L\times SU(2)_R$ isometry.
We will next perform a non-abelian transformation with respect to one of these $SU(2)$
factors. As we shall see, the two cases differ by a simple sign factor in the NS three-form.

\no
In order to perform the T-duality transformation we will need the Hodge-dual of the above three-form
\be
F_7 = -(\star F_3) = \left({\rm Vol}(S^3)+ {\rm Vol}(AdS_3)\right)\wedge {\rm Vol}(T^4) \ ,
\ee
where the sign in the definition has been chosen in accordance to \eqn{hdhi2}.

\subsection{ The T-dual on the NS sector}

We now apply our general reasoning for PCMs to the case of interesting where the group $G=SU(2)$.
The representation
matrices and algebra structure constants are given by
\be
t_i = {\s_i\ov\sqrt{2}}\ ,\qq f_{ijk} = \sqrt{2} \e_{ijk} \ ,
\ee
where $\s_i$ are the standard Pauli-matrices.
Setting $v_i =  x_i/\sqrt{2}$ we have that
\be
M_{ij}= \d_{ij} + \e_{ijk} x_k \quad  \Longrightarrow \quad
(M^{-1})^{ij} = {1\ov 1+ r^2}(\d_{ij} + x_i x_j - \e_{ijk} x_k)\ .
\ee
We note that had we gauged the right $G_R$ isometry would have amount to flipping the sign of the
Lagrange multiplier $v_i$. For the case of the example of interest we should send $x_i\to  - x_i$.

\no
This dual $\s$-model has by construction a residual $SU(2)$ symmetry that can be made
manifest by introducing spherical coordinates in place of the Cartesian ones.
We obtain that the fields of the NS sector of dual model are
\ba
&& ds^2 = ds^2({\rm AdS_3})  + dr^2  + {r^2\ov 1+ r^2} d\Om_2^2 + ds^2(T^4)\ ,
\nonumber\\
&& B= \e {r^3\ov 1+r^2} {\rm Vol}(S^2) \quad \Longrightarrow  \quad
H = \e {r^2 (3 + r^2)\ov (1 + r^2)^2}  \ dr\wedge {\rm Vol}(S^2)\ ,
\label{nsnsn}
\\
&& \Phi  = -\ha\ln(1+r^2)\ ,\qq  \e=\pm 1\ .
\nonumber
\ea
The two signs parametrized by the arithmetic parameter $\e$ correspond to gauging the right (for $\e=1$)
and left (for $\e=-1$) $SU(2)$ isometries, respectively.
The above background corresponds to a smooth space, due to the fact that the isometry acts with no isotropy,
and interpolates between $\mathbb{R}^3 $ and $\mathbb{R} \times S^2$
(times the $AdS_3\times T^4$ part).\footnote{The part of the above metric and antisymmetric
tensor associated to the non-abelian of PCM for $S^3$ are essentially the same as
those computed in \cite{Fridling:1983ha,Fradkin:1984ai}.}

\no
In what follows some calculations are most easily performed
by introducing vielbeins that allow to work with tangent frame Gamma matrices.
For the dualised directions we have a convenient choice of three one-forms
\be
e^i = {1\ov \sqrt{1+r^2}} (dx^i +  x^i a(r) dr)\ ,\qq a(r)= {\sqrt{1+r^2}-1\ov r}\ .
\label{vb}
\ee
%such that
%\be
%\sum_{i=1}^3 e^{i }\otimes e^{i} =    dr^2  + {r^2\ov 1+ r^2} d\Om_2^2\, .
%\ee

\subsection{The T-duality transformation on the RR-fluxes}

%\subsubsection{The construction}

\no
The Lorentz transformation matrix \eqn{Lambda} is, in this case, explicitly given by
\be
\L_i{}^j = -(MM^{-1T})_i{}^j = {r^2-1\ov r^2+1}\ \d_{ij}
- {2\ov r^2+1} (x_i x_j +\e_{ijk}x_k)\ ,
\ee
with $r^2=x_ix_i$ and where we have used
\be
M_{ij} = \d_{ij} + \e_{ijk}x_k \ ,
\qq M^{-1}_{ij} = {1\ov r^2+1}( \d_{ij} + x_ix_j- \e_{ijk}x_k)\ .
\ee

\no
The matrix $\Om$ that solves \eqn{spino1} is given by
\be
\Om = \G_{11} \tilde \Om \ ,\qq \tilde \Om = {\G_{123} + {\bf x\cdot \G}\ov \sqrt{1+r^2}} \ .
\ee
The matrix $\tilde \Om$ is unitary satisfying \eqn{spino1}
but with a minus in the right hand side.
Restricting first our attention to the directions being dualised,  we take the Gamma matrices
$\G_{i}$ for $i=1,2,3$ to obey
\be
\G_i \G_j = \d_{ij} + i \e_{ijk}\S_k\ ,
\ee
 and
 \ba
&& \S_i \S_j = \d_{ij} + i\e_{ijk}\S_k\  ,
\nonumber\\
&& \G_i \S_j = - i \G_{123} \d_{ij} + i\e_{ijk}\G_k\  ,
\\
&& \S_i \G_j =-i \G_{123} \d_{ij} + i\e_{ijk}\G_k\  ,
\nonumber
\ea
where $\G_{123}=\G_1\G_2\G_3$, obeying
\be
\G_{123}^2 = -I\ , \quad \G_{123} \G_i = \G_i \G_{123}= i \S_i \ ,\quad
\G_{123} \S_i = \S_i \G_{123}= i \G_i\ .
\ee
The set of matrices $\{\G_i,\S_i\}$ form an $SO(4)$ algebra.
A useful representation in computations is
\be
\G_i = \s_3\otimes \s_i \ ,\qq \S_i = I_2\otimes \s_i \ ,\qq \G_{123} = i \s_3\otimes I\ .
\ee
where $\s_i$ are the usual $2\times 2$ Pauli matrices.\footnote{Of course, one should be more exact
and use a representation for these $\Gamma$ matrices corresponding
to the full ten dimensional Clifford algebra. However this abbreviated representation is sufficient to demonstrate the result.}
Using this representation for the Gamma matrices  we immediately
see that
\be
\tilde\Om =\diag(\om,-\om)\ ,\quad \om = {i I +  {\bf x}\cdot  \s\ov \sqrt{1+r^2}} \  .
\ee
Then we compute
\be
\om^{-1} \s^i \om = {1-r^2\ov r^2+1}\ \s_i +{2\ov r^2+1}
\left[ ({\bf x}\cdot \s)x_i  + \e_{ijk} x_j \s_k \right] = - \L^i{}_j \s^j \ ,
\label{dj2}
\ee
thus proving our assertion.  To verify the equation \eqn{spino1}
for the the remaining transverse direction with $a\neq 1,2,3$, one simply uses $\{\G_{11},\G^a\}=\{\G^i, \G^a\}=0$.

\subsubsection{The transformation}

We may now apply the transformation rule for RR flux \eqn{ppom}.
We have that
\be
 P= \ha \left(\frac{1}{3!} \slashed{F}_{3} + \frac{1}{7!} \slashed{F}_{7}\right) =
 %\ha\left(\G^{0'1'2'} + \G^{123} + \G^{0'1'2'4567}+ \G^{1234567} \right)\ ,
\ha\left(\G^{0'1'2'} + \G^{123}\right)\left(\mathbb{1} + \G^{4567} \right)\ ,
\ee
where the indices $0'1'2'$ correspond to the $AdS_3$ directions, $123$ to $S^3$ and $4567$ to $T^4$.
Given this and the structure of $\Omega$ we expect for $\hat P$ the form
\be
\hat P = {e^{\Phi}\ov 2}\left(I F_0 +{1\ov 2}
{\slashed F}_{2} + {1\ov 4!} {\slashed F}_{4} + {1\ov 6!} {\slashed F}_{6} + \frac{1}{10!}
\slashed{F}_{10}\right)\ .
\ee
After some rearrangements we find that
\be
F_0 = 1\ ,
\label{mmm}
\ee
and that
\ba
&& F_{10} = -{\rm Vol}(G) = -{\rm Vol}(AdS_3)\wedge {\rm Vol}(S^3)\wedge  {\rm Vol}(T^4)\ ,
\nonumber\\
&& F_6 = {r^2\ov 1+r^2} \left[r {\rm Vol}(T^4) + dr\wedge {\rm Vol}(AdS_3)\right]\wedge {\rm Vol}(S^2)\ .
\ea
For the 2-form we find
\ba
 F_2 &= &  x_1 e^2 \wedge e^3 +  x_2 e^3 \wedge e^1 +  x_3 e^1 \wedge e^2
\nonumber\\
&= & {1\ov 1+r^2}(x_1 dx_2 \wedge dx_3 + x_2 dx_3 \wedge dx_1 + x_3 dx_1 \wedge dx_2)
\label{mmm1}
\\
& = &
{r^3\ov 1+r^2}\ {\rm Vol}(S^2) \ .
\nonumber
\ea
Finally, we find for the 4-form and the 8-form
\be
F_4 = {\rm Vol}(AdS_3) \wedge x_i e^i + {\rm Vol}(T^4)= - r dr\wedge {\rm Vol}(AdS_3) + {\rm Vol}(T^4) \ ,
\label{mmm2}
\ee
and
\be
 F_8 = {\rm Vol}(AdS_3) \wedge x_i e^i \wedge {\rm Vol}(T^4)= - r dr\wedge {\rm Vol}(AdS_3) \wedge {\rm Vol}(T^4) \ .
\label{mmm22}
\ee
One can easily verify that $ \star F_2= F_8 $, $\star F_4= - F_6 $ and $\star F_{10} = 1$,
as expected from \eqn{hdhi1}.

 %%%%%%%%

\subsection{Verifying the field equations of massive IIA supergravity}
We have now obtained a dual background
given by \eqn{nsnsn} supported by the RR-fluxes
\eqn{mmm}, \eqn{mmm1} and \eqn{mmm2}.
We would like to verify that indeed it solves the equations of field equations
of the massive type-IIA supergravity. 

To show this we shall assume the specific form of the NS fields in \eqn{nsnsn} and
for the fluxes we try an ansatz of the form
\be
F_2 = A(r) {\rm Vol}(S^2)\ ,\qq
 F_4  = B(r) dr\wedge {\rm Vol}(AdS_3) + C(r) {\rm Vol}(T_4)\ ,
\ee
which preserves the symmetries of the NS background fields. The functions
$A(r)$, $B(r)$ and $C(r)$ will be determined by satisfying the Bianchi identities and
flux equations.  From the Bianchi identity we immediately find that
\be
A(r) = m\e {r^3\ov 1 + r^2} + c_1\ , \qq C(r)= c_2\ ,
\ee
where $c_1$ and $c_2$ are constants.  The latter constant must be
non-zero to admit a solution, but can be set  $c_2 = \pm 1$ without loss of generality.
In fact, there is a flip symmetry that allows us to take $c_2= 1$
(this is not symmetry of the flux equations \eqn{fluxx} in general, but is specific our ansatz).

\no
Next we compute the Hodge duals
\ba
&& \star H = -{h(r)\ov f(r)} {\rm Vol}(AdS_3)\wedge {\rm Vol}(T^4)\ ,
\nonumber\\
&&\star F_2 = {A(r)\ov f(r)} {\rm Vol}(AdS_3)\wedge dr\wedge  {\rm Vol}(T^4)\ ,
\\
&& \star F_4 = B(r) f(r) {\rm Vol}(S^2)\wedge {\rm Vol}(T^4)
+ f(r) {\rm Vol}(AdS_3)\wedge dr\wedge  {\rm Vol}(S^2)\ ,
\nonumber
\ea
where
\be
f(r) = {r^2\ov 1+r^2}\ ,\qq h(r)= \e {r^2 (3 + r^2)\ov (1 + r^2)^2} \ .
\ee
Then the flux equations give the conditions
\be
{d\ov dr}(B(r) f(r)) =  - h(r)\ ,\qq {d\ov dr}\left(e^{-2\Phi}{h(r)\ov f(r)}\right) +  B(r)
= m{A(r)\ov f(r)}\ .
\ee
Solving them requires that
\be
m=\pm 1  \ ,\qq B(r) = -\e\ r + c_1 {1+r^2\ov r^2}\ .
\ee
It can be readily verified that the dilaton equation \eqn{dilaeq} is satisfied.
It remains to check Einstein's equations \eqn{Einsteq}. It turns out that these
are satisfied provided that the constant $c_1=0$. Hence, the RR-fluxes assume the form
\be
F_2 = m\e {r^3\ov 1+r^2}\ {\rm Vol}(S^2)\ ,\qq
 F_4  = - \e r dr\wedge {\rm Vol}(AdS_3) +  {\rm Vol}(T_4)\ .
\label{form24}
\ee
We see that this matches the background found by the T-duality transformation if $\e=1$ and $m= 1$.

\no
We also note that in the small $r$ limit the spacetime is that for $AdS_3\times R^3 \times T^4$,
whereas for $r\to \infty$, that for $AdS_3\times R\times S^2 \times T^4$.
However, these limiting behaviors cannot be promoted to full supergravity solutions on their own,
since, as it turns out, keeping the leading order behaviour of the fluxes and the dilaton is not
consistent with the equations of motion.

\subsection{Supersymmetry of the T-dual background}

Bosonic backgrounds will preserve a fraction of supersymmetry if the
dilatino and gravitino variations with all fermions set to zero admit non-trivial spinors as solutions.
These variations are given by
\be
\delta \lambda = \left( \slashed{\partial}\Phi + \frac{1}{2 \cdot 3!}
\slashed{H} \Gamma_{11}  \right)\varepsilon  +\frac{1}{4} e^{\Phi} \sum_{n=0}^2
\frac{5-2n}{(2n)!}\slashed{F}_{2n} (\Gamma_{11})^{n} \varepsilon \ ,
\label{form25d}
\ee
and
\be
 \delta \psi_{\mu} = D_{\mu} \varepsilon  - \Lambda_{\mu} \varepsilon
\equiv \left( \partial_{\m} + \frac{1}{4} \slashed{\omega}_{\m}
+\frac{1}{8}\G_{11}\slashed{H}_{\m}\right) \varepsilon
+ \frac{e^{\Phi}}{8}  \sum_{n=0}^2 \frac{1}{(2n)!}\slashed{F}_{2n}\G_{\m} (\Gamma_{11})^{n} \varepsilon  \ ,
\label{form25}
\ee
where $D_\m$ is the usual covariant derivative built  with the spin connection.
Let us consider whether the T-dual background given by \eqn{nsnsn}, \eqn{mmm}, \eqn{mmm1} and \eqn{mmm2}
preserves a fraction of supersymmetric of the original
background which is $\ha$-supersymmetric, i.e. it preserves sixteen real supercharrges. We will
explicitly examine and solve the above
dilation and gravitino supersymmetry variations.
Since the dual geometry
has all fluxes turned on it may seem at fist sight that this is an unlikely proposition.
However, for spinors obeying the projector conditions
\be
\label{proj1}
\G^{0'1'2'123}\G_{11}\varepsilon = \varepsilon
\ee
and
\be
\frac{1}{\sqrt{1+r^{2}}}\left(r \G^{1} - \G^{123}\G_{11}\right)\varepsilon =  \varepsilon\ ,
\label{proj2}
\ee
 one can establish that the dilatino variation identically vanishes.
 The first projector is simply inherited from the original $AdS_{3} \times S^{3}\times T^4$
 geometry whereas the second projector is more exotic.    This position dependent projector can
  be thought of as being related to the Lorentz frame rotation induced by performing the
   T-duality.\footnote{We note that similar position depend projectors arise
   in considering  $\kappa$-symmetry and branes intersecting at angles  \cite{Bergshoeff:1997kr}. In
searching for supersymmetry preserving supergravity solutions similar examples were found
in \cite{Edelstein:2002zy} . }
    A spinor obeying these commuting projectors possesses eight real
    degrees of freedom indicating a $\frac{1}{4}$-BPS solution.

\no
  To complete this analysis one must also ensure  that the gravitino equation vanishes
  for   spinors obeying the projectors and which are of a suitable functional form.  By differentiating the
   gravitino equation one can form an integrability condition
 \be
 \frac{1}{4} R_{\mu \nu ab}\G^{ab} \varepsilon = \left( D_{\mu}\Lambda_{\n} -
  D_{\nu}\Lambda_{\m} - [\L_{\m}, \L_{\n}] \right) \varepsilon \, .
 \ee
   For our particular background and for spinors obeying   the above projector
   conditions one  can verify that this integrability condition indeed holds thereby ensuring supersymmetry.
   Indeed, one can also integrate the gravitino equation to give an explicit form for the Killing spinors
  \be
  \varepsilon = \Omega_{\rm AdS}\cdot\Omega_{r} \cdot \Omega_{\th}\cdot \Omega_{\phi}\cdot \varepsilon_{0}\ ,
  \ee
  where $\varepsilon_{0}$ is a spinor obeying the projectors  \eqn{proj1} and \eqn{proj2} and where
 \ba
 \Omega_{r} &=& \exp\left(  \ha \tan^{-1}(r) \G^{23}  \right)\, ,\qq
   \Omega_{\phi}= \exp\left( \frac{\phi}{2} \G^{23}  \right)\ ,
\nonumber
\\
  \Omega_{\th} &=& \exp\left( -\frac{\th}{2(1+r^{2)}}
\left[ -\G^{12}  + r^{2}\ \G^{12}\Gamma_{11}  -r \ \G^{13} - r \G^{13}\G_{11}   \right] \right)\ .
    \ea
The factor $\Omega_{\rm AdS}$ contains the functional dependance on $AdS_{3}$
and depends on the coordinate system of our preference
(in horospherical coordinate system it is given by \cite{Lu:1998nu}).
Note that although the four-form does have legs in the $T^4$,
the projection conditions obeyed by $\varepsilon_{0}$ ensure that the gravitino
equation in these direction simply boils down to ensuring that the spinor is
 constant with respect to these coordinates (as would be expected from the geometry).

\subsubsection{Remarks on the Kosmann derivative}

 We have seen by an explicit computation
 that the dual geometry preserves eight supersymmetries
 whereas the  $AdS_{3}\times S^{3} \times T^{4}$ solution of type-IIB supergravity
 that we started with enjoys sixteen supersymmetries. Hence half of the
  supersymmetries have been destroyed through the dualisation
procedure. This is a statement valid in the supergravity low energy
approximation of string theory. Stringy winding modes may
reestablish supersymmetry \cite{Bakas:1995hc, Duff:1997qz}. The origin of
this supersymmetry breaking
can also be seen by considering how the $SU(2)$ symmetry
along which we dualise  acts on the Killing spinors of the original
background.

\no
In general, it is ambiguous to define the action of a vector on a spinor,
however the action of a Killing vector, $k = k^{\m}\partial_{\m}$,
on a spinor is well defined and is given by the Kosmann (spinor-Lorentz-Lie) derivative \cite{Kosmann}
\be
{\cal L}_k \e = k^\m D_\m \e  - \frac{1}{4} \nabla_\m k_\n \G^{\m\n} \e \ .
\ee
This derivation maps spinors to spinors and induces on bispinors the
 usual action of the Lie derivative.
 Additionally, it forms representations of the algebra of vector fields so that
\be
[{\cal L}_{k_1} ,{\cal L}_{k_2}]  \e = {\cal L}_{[k_1, k_2]} \e \
\ee
and for further properties the reader may consult \cite{oFarrill:1999va,Ortin:2002qb}.
We may thus ask how many of the Killing spinors of the $AdS_{3}\times S^{3} \times T^{4}$
background are  invariant under the $SU(2)_L$ symmetry in the sense that their Kosman derivative vanishes.
In this case the r\^ole of the Killing vector $k$ above is played by the three vectors $K^a$, $a=1,2,3$,
with components $K^a_i$ and inverse ones $K^i_a$.
These obey the useful identities
  \ba
 && \sum_{a=1}^3 K^{a \ i}K^a_j =  \d_j^i   \   , \qq
  \sum_{i=1}^3 K^{a \ i}K^b_i = \delta^{ab}\ ,\qq
 K^a_iK^b_j K^c_k \e_{abc} = - \e_{ijk} \ ,
\ea
as well as derivative relations
\be
\nabla_i K_j^a + \nabla_j K_i^a = 0 \ , \qq
\nabla_i K_j^a - \nabla_j K_i^a = -\e_{abc}K^b_i K^c_j \  ,
\ee
which are just the Killing property and the Maurer--Cartan equations, respectively.
In the above $\e_{ijk}$ is a density containing ${\rm Vol}(S^3)$.

\no
The sixteen independent Killing spinors of  type-IIB supergravity
obeying a projection condition arising from the dilatino variation
$\slashed{F}_{3} \e = 0$,
which using \eqn{adsfl} implies that
\be
\G^{0'1'2'123} \e = -\e \ .
\ee
They also satisfy
\be
D_\m \e -{i\ov 8\cdot 3!} \slashed F_3 \G_\m \e^* = 0 \ ,
\ee
arising form the gravitino variation. This can be written in the convenient form
\be
( \mathbb{1}_2  \otimes D_\mu  ) \varepsilon  -   \frac{1}{8 \cdot 3!}
 (  \s_1  \otimes \slashed{F}_3\G_\mu) \varepsilon= 0\ , \qq
\varepsilon =   \left(\begin{array}{c}
    \e_1 \\
    \e_2
    \end{array} \right)\ ,
\ee
where we have reassembled the complex chiral spinor $\e = \e_1 + i \e_2$ into a doublet.
Using the above notation and variation we may write the Kosman derivative as
\ba
{\cal L}_a   \varepsilon & =& (\mathbb{1}_2 \otimes K^i_a D_i)  \varepsilon
- \frac{1}{4} (\mathbb{1}_2 \otimes  \nabla_i K^a_j \G^{ij})  \varepsilon
 \nonumber \\
&=&  \frac{1}{8 \cdot 3!}   (  \s_1  \otimes K^i_a  \slashed{F}_3\G_i)\varepsilon
- \frac{1}{4} (\mathbb{1}_2 \otimes  \nabla_i K^a_j \G^{ij})  \varepsilon\ .
\ea
Then, using for the first term that
\be
{1\ov 3!} \slashed F_3 \G_i \varepsilon
= (\G^{0'1'2'}+\G^{123})\G_i \varepsilon = 2 \G_i \G^{123}\varepsilon
= \e_{ijk}\G^{jk}\
\ee
and for the second the above Maurer--Cartan equations, we obtain that
\be
{\cal L}_a   \varepsilon = -{1\ov 4}
\left(  {\cal P}  \otimes  k_a^i \e_{ijk} \G^{jk}  \right) \varepsilon\ , \qq
{\cal P} = \ha(\mathbb{1}_2 -\s_1)\ .
\ee
Demanding that this vanishes leads to a condition on $\varepsilon$ which
can be simplified by squaring it, contracting appropriately over the index $a$
and by using the fact that $\cal P$ is a projector.
We find a necessary condition for the Killing spinor to be invariant is that
\be
({\cal P}   \otimes \mathbb{1}_{32})  \varepsilon = 0\  .
\ee
It can be easily seen that this is also sufficient.
In conclusion, exactly half of the supersymmetries  commute (in the above sense)
with the $SU(2)$ action and we should only expect half the original supersymmetry to be preserved
in the T-dual geometry. This is in exact accordance with the explicit result.

\section{Non-abelian T-dual in the D3 near horizon}

Our second example concerns the type-IIB supergravity solution describing the near horizon limit
of the D3-brane background. It consists of a metric
\be
ds^2 = ds^2({\rm AdS_5}) + ds^2({\rm S^5}) \ ,
\ee
normalized such that $R_{\m\n}= \mp g_{\m\n}$ for the $AdS_5$ and $S^5$ factors, respectively,
supported by the self-dual Ramond flux
\be
F_5 = 2 {\rm Vol}(AdS_5) -2  {\rm Vol}(S^5)\ .
\ee
As before we note that we have completely absorbed all constant factors by appropriate rescalings.
We would like to perform a non-abelian T-duality with respect to an $SU(2)\in SO(6)$ symmetry.
For that purpose we write the line element for $S^5$ in the form
\be
 ds^2({\rm S^5}) = 4(d\th^2 + \sin^2\th\ d\phi^2) + \cos^2\th\  ds^2({\rm S^3})\ ,
\ee
where for the $S^3$ factor the normalization is such that $R_{\m\n}=\ha g_{\m\n}$.

\subsection{ The construction of the T-dual background}

To obtain the dual background we follow the same steps as before.  In particular,
we find that the matrix $M_{ij}$ occurring in the dual $\sigma$-model \eqn{vKv} is given by
\be
M_{ij} = \cos^2\th\ \d_{ij} +\e_{ijk}\ x_k \ \Longrightarrow\
M^{-1}_{ij} = {\cos^2\th\ \d_{ij} + {\cos^{-2}\th}\ x_i x_j - \e_{ijk}\ x_k\ov \cos^4\th + r^2}\ .
\ee
Finally, we obtain that the fields of the NS sector of dual model are
\ba
\label{dualNS}
&& ds^2 = ds^2({\rm AdS_5})  +4(d\th^2 + \sin^2\th d\phi^2) +
{dr^2\ov \cos^2\th}  + {r^2 \cos^2\th \ov \cos^4\th+ r^2}d\Om_2^2 \ ,
\nonumber\\
&& B=  {\e r^3\ov \cos^4\th +r^2} {\rm Vol}(S^2)\ , \quad \nonumber\\
&&
H ={\e r^2\ov  (\cos^4\th + r^2)^2} \left[(3\cos^4\th + r^2)dr +
4 r \sin\th \cos^3\th\ d\th\right] \wedge {\rm Vol}(S^2)\ ,
\\
&& \Phi  = -\ha\ln [\cos^2\th (\cos^4\th+r^2)]\ ,\qq  \e=\pm 1\ ,
\nonumber
\ea
where once again the arithmetic parameter $\e$ indicates the two distinct possibilities in performing
the non-abelian T-duality transformation.
This background has an $SO(2,4)\times SU(2) \times U(1)$ symmetry.
We will comment more later on this point in relation to the amount of
supersymmetry preserved by the solution and its eleven dimensional supergravity lift.

\no
We now compute the fluxes that should support the geometry and make into
a full solution of type-IIA supergravity. In this case there is no
zero-form produced since there is no three-form or one-form flux in the original background.
We will use again \eqn{ppom}, but now with
\be
 \Om ={ \cos^2\th \ \G_{123} +  {\bf x}\cdot {\bf \G}\ov \sqrt{\cos^4\th +r^2}}
\ee
and
\ba
&& P= \G^{0'1'2'3'4'} - \G^{12345}\ ,
\nonumber\\
&& \hat P =  {e^{\Phi}\ov 2}\left({1\ov 2}
{\slashed F}_{2} + {1\ov 4!} {\slashed F}_{4} + {1\ov 6!} {\slashed F}_{6} +
{1\ov 8!} {\slashed F}_{8}\right)\ ,
\ea
where the indices $0'1'\cdots 4'$ correspond to the $AdS_5$ factor; $1$ to $r$; $2,3$
to the coordinates in $S^2$; and $4,5$ to $\th$ and $\phi$.
Converting all Gamma matrices to tangent frame using the drei-bein
\be
e^i = {1\ov \cos\th \sqrt{\cos^4\th +r^2}} (\cos^2\th\ dx^i +  x^i b(r) dr)\ ,\quad
b(r)= {\sqrt{\cos^4\th +r^2}-\cos^2\th\ov r}\ ,
\ee
we find that
\ba
\label{dualRR}
&& F_8 =- 2 {r^2\cos^4\th\ov \cos^4\th + r^2}\ {\rm Vol}(AdS_5) \wedge dr\wedge {\rm Vol}(S^2)\ ,
\nonumber
\\
&& F_6 =-2 {r}\ dr\wedge  {\rm Vol}(AdS_5)\ ,
\nonumber
\\
&& F_4 = - 8 {r^3 \cos^3\th \sin\th \ov \cos^4\th + r^2} \ d\th \wedge d\phi \wedge  {\rm Vol}(S^2)\ ,
\\
&& F_2 = -8 \cos^3\th \sin\th\ d\th\wedge d\phi \ .
\nonumber
\ea
Note that $\star F_6=F_4$ and $\star F_8=-F_2$ in accordance with \eqn{hdhi1}.
It is easy to check that the Bianchi
identities \eqn{biiac} (with $m=0$) are obeyed provided we choose $\e=1$.
In addition, one can readily see that the equations of
motion \eqn{fluxx} for the fluxes (again with $m=0$) are indeed satisfied.
For the $H$-equation one uses that
\be
e^{-2\Phi}\star H = -4 \sin\th \cos\th \left(-r \sin\th \cos\th \ dr + (3\cos^4\th + r^2)d\th\right)
 \wedge d\phi \wedge {\rm Vol}(AdS_5)\ ,
\ee
from which we find
\be
 d\left(e^{-2\Phi}\star H \right)
= - 16 r \cos^3\th \sin\th\ dr\wedge d\th \wedge d\phi \wedge {\rm Vol}(AdS_5)\ .
\ee
We have also checked that the Einstein's and dilaton
equations \eqn{Einsteq} and \eqn{dilaeq} are satisfied.

\subsection{Supersymmetry of the T-dual background}

We first examine the dilatino equation in \eqn{form25d}. It is convenient
for notational purposes to write $r=R\cos^2\th$. Using the flat index notation we have
\ba
&& H_{123}= {R^2+3\ov \cos\th(R^2+1)}\ ,\qq H_{234}= {2 R \sin\th \ov \cos\th(R^2+1)}\ ,
\nonumber\\
&& (F_2)_{45}=-2 \cos^3\th\ ,\qq (F_4)_{2345}=-2 R \cos^3\th\ .
\ea
Then, after
multiplying \eqn{form25d} by $2 \cos\th (R^2+1)/(R^2+3)$, the dilatino equation can be cast into the form
\be
\G_{123}\G_{11}\varepsilon = (A+B)\varepsilon \ ,
\label{hhg}
\ee
where the matrices
\ba
& & A = {R\sqrt{R^2+1}\ov R^2+3} \cos\th\ \G_{2345} -\sin\th\ \G_4\ ,
\nonumber\\
&& B = 3 {\sqrt{R^2+1}\ov R^2+3}\cos\th\ \G_{45} \G_{11}
+  {2R\ov R^2+3}\left( \G_1- \sin\th \G_{234}\G_{11}\right)\ .
\label{hhg1}
\ea
We would like to write \eqn{hhg} in the form
\be
{\cal P}\e = \e\ ,\qq {\cal P}^2 = \mathbb{1}\ .
\label{fhhpo}
\ee
We note the relations $(\G_{123}\G_{11})^2=\mathbb{1}$ and $[\G_{123}\G_{11}, A]=\{\G_{123}\G_{11}, B\}=0$.
Using them and acting on \eqn{hhg} with $\G_{123}\G_{11}$ we obtain
\be
(A^2-B^2+[A,B])\varepsilon = \varepsilon\ .
\label{jfj1}
\ee
Then we compute
\ba
&& A^2 = {R^2(R^2+1)\ov (R^2+3)^2} \cos^2\th + \sin^2\th\ ,
\nonumber\\
&& B^2= {1\ov (R^2+3)^2}\left[-9(R^2+1)\cos^2\th + 4 R^2(1+\sin^2\th)\right]
\nonumber\\
&&\phantom{xxxxxxxx}
 -4  {R\ov (R^2+3)^2}\sin\th \left(3 \sqrt{R^2+1} \cos\th\ \G_{235} + 2 R \G_{1234}\G_{11}\right)\ .
\\
&& [A,B] = - 4 {R^2\sqrt{R^2+1}\ov (R^2+3)^2}\sin\th \cos\th \ \G_5\G_{11}
+ 4 {R\ov R^2+3} \sin\th  (\sin\th\ \G_{23}\G_{11} + \G_{14})\ .
\nonumber
\ea
We further simplify \eqn{jfj1} by using \eqn{hhg} to write
$\G_{1234}\G_{11}\varepsilon= -\G_4(A+B)\varepsilon$. The result can be cast in the form \eqn{fhhpo} with
\be
{\cal P} = {\cos\th \ov \sqrt{r^2+\cos^4\th}}(\cos^2\th\ \G_{12345}-r \G_{145}\G_{11})
+ \sin\th\ \G_{1234}\G_{11}\ ,
\ee
where we have restored the original $r$ variable.
One may readily verify after some algebraic manipulations that indeed ${\cal P}^2= \mathbb{1}$.

\no
Hence we have seen that the dilatino equation is solved by supersymmetry
parameters obeying a single, albeit somewhat complicated, field-dependent projector condition,
projecting out half of the components leaving a possible of
sixteen surviving supersymmetries.  The complicated nature of the background
 makes it rather difficult to give an explicit formula for the Killing spinors in this case.
 However, one can check, for instance using the {\tt Mathematica} computer programme,
that for such a restricted spinor the integrability condition for the gravitino equation is obeyed.
 Hence this T-dual background is found to be supersymmetric, preserving precisely sixteen supercharges.

\no
As before we may also infer the amount of supersymmetry that
 will be preserved after the T-duality is performed
by examining the action of the Kosmann derivative on the Killing
spinor of the original background for $AdS_5 \times S^5$.
This satisfies
\be
D_\m\varepsilon -{i\ov 4\cdot 5!} \slashed F_5 \G_\m \e = 0 \ ,
\ee
arising form the gravitino variation. There is of course
no associated projection since the dilatino equations is identically satisfied, i.e.
the background is maximally supersymmetric. The above Killing spinor equation
 can be written in the convenient form
\be
( \mathbb{1}_2  \otimes D_\mu  ) \varepsilon  +   \frac{1}{4\cdot 5!}
 ( i \s_2  \otimes \slashed{F}_5\G_\mu) \varepsilon= 0\ , \qq
\varepsilon =   \left(\begin{array}{c}
    \e_1 \\
    \e_2
    \end{array} \right)\ ,
\ee
where as before we have assembled the two chiral spinors $\e_i$ into a doublet.
Using the above notation and variation we may write the Kosmann derivative as
\ba
{\cal L}_a   \varepsilon & =& (\mathbb{1}_2 \otimes K^\m_a D_\m)  \varepsilon
- \frac{1}{4} (\mathbb{1}_2 \otimes  \nabla_\m K^a_\n \G^{\m\n})  \varepsilon
 \nonumber \\
&=&
 -\frac{1}{4\cdot 5! }   ( i \s_2  \otimes K^\m_a  \slashed{F}_5 \G_\m)\varepsilon
- \frac{1}{4} (\mathbb{1}_2 \otimes  \nabla_\m K^a_\n \G^{\m\n})  \varepsilon\ .
\ea
%\be
% \slashed F_3 \G_i \varepsilon
%= (\G^{0'1'2'}+\G^{123})\G_i \varepsilon = 2 \G_i \G^{123}\varepsilon
%={\rm Vol}(S^3) \e_{ijk}\G^{ijk}\
%\ee
%and for the second the above Maurer--Cartan equations we obtain that
%\be
%{\cal L}_a   \varepsilon = {1\ov 8} {\rm Vol}(S^3)
%\left[ (  \s_1  \otimes  k_a^i \e_{ijk} \g^{jk}  )  -  (\mathbb{1}_2  \otimes  k_a^i \e_{ijk} \g^{jk}  )
%\right] \varepsilon\ .
%\ee
This is quite complicated to analyze analytically in full detail, the reason being that
the three-dimensional part obtained by T-duality is coupled to the rest of the
coordinates, in particular $\th$. Nevertheless,
we have verified, on computer, that demanding vanishing Kosmann derivative
places constraints on the spinor $\varepsilon$ which project out exactly half of the degrees of freedom.
Hence from this perspective we also recover the expectation that the dual background should
be $\ha$-BPS, which, of course, has been explicitly shown above.

\subsection{M-theory lift and its interpretation}

The solution found above in \eqn{dualNS} and \eqn{dualRR} has a natural eleven-dimensional origin.
In  \cite{Gaiotto:2009gz}  the gravity duals for a large class
of generalized quiver $\cN =2 $ superconformal field theories \cite{Gaiotto:2009we}
were presented.  These eleven dimensional geometries contain $AdS_5$ factors,
they possess $SO(2,4)\times SU(2)\times U(1)$ isometry, as required by $\cN=2$ superconformal
invariance and fall into the general
ansatz of \cite{Lin:2004nb} (for related earlier work see also \cite{Gauntlett:2004zh}).
In general these geometries depend on a single function
satisfying the continual Toda equation \cite{Saveliev}.
However, if this solution has an additional
$U(1)$ symmetry then the geometry simplifies and takes the form
\begin{eqnarray}
\label{11metric}
ds^2_{11}&=& \left(\frac{\dot{V}\tilde{\Delta}}{2V''}\right)^{1/3}
\left[d s^2_{AdS_5} +\frac{2V''\dot{V}}{\tilde{\Delta}} d\Om^2_2 + \frac{2V''}{\dot{V}}\left(d\rho^2+
\frac{2\dot{V}}{2\dot{V}-\ddot{V}}\rho^2 d\phi^2
+d\eta^2\right)\right. \nonumber\\
&&\left.\qquad\qquad\qquad\qquad+\frac{2(2\dot{V}-\ddot{V})}{\dot{V}\tilde{\Delta}}
\left(d\beta+ \frac{2\dot{V}\dot{V}'}{2\dot{V}-\ddot{V}}d\phi\right)^2 \right]\ ,
\nonumber\\
C_3&=&2 \left[-2\frac{\dot{V}^2V''}{\tilde{\Delta}}d\phi+
\left(\frac{\dot{V}\dot{V}'}{\tilde{\Delta}}-\eta\right)d\beta\right]\wedge
{\rm Vol}(S_2)\ ,
\end{eqnarray}
where
\be
\tilde{\Delta} =(2\dot{V}-\ddot{V})V''+(\dot{V}')^2 \
\ee
and where we used the definitions
\be
\r \del_\r V= \dot V \ ,\qq \del_\eta V =V'\ \, .
\ee
The function $V=V(\r,\eta)$ is a rotational invariant solution of the Poisson equation in
three dimensions with cylindrical polar coordinates $\r,\phi,\eta$, i.e.
\be
\ddot V + \r^2 V'' = \lambda(\eta) \r \d(\r) \ .
\label{lapls}
\ee
By Gauss' law, equivalently by dividing this equation by $\r$ and then integrating,
one can determine that the charge density is given by
\be
\lambda(\eta) = \dot{V} (\eta, \rho = 0 )\  .
\ee
For physically acceptable backgrounds the basic requirement for the
line density is that it should be composed of segments linear in $\eta$ with integer slopes.

\no
Since the background \eqn{11metric} is also isometric around
the periodic $\beta$ direction one can perform a Kaluza-Klien reduction to arrive at
a ten-dimensional type-IIA geometry as detailed in \cite{ReidEdwards:2010qs}.
The result for the metric is
\be
\label{10metric}
ds^2_{10}=\left(\frac{2\dot{V}-\ddot{V}}{V''}\right)^{\frac{1}{2}}
\left(ds^2_{AdS_5} +\frac{2V''\dot{V}}{\tilde{\Delta}}d\Om^2_2 +
\frac{2V''}{\dot{V}}\left(d\rho^2+d\eta^2\right)+
\frac{4V''}{2\dot{V}-\ddot{V}}\rho^2 d\phi^2\right)\,.
\ee
By making the coordinate transformations
\be
 \r =\sin \th\ ,\qq   \eta = \frac{r}{2}\ ,
\ee
we can bring the solution found in  \eqn{dualNS} and  \eqn{dualRR} into
exactly this form with the potential given by
\be
\label{ourpot}
V = \eta \ln \r + \eta \left( {\eta^2\ov 3}-   {\r^2\ov 2}  \right)   \ .
\label{vvv}
\ee
This potential satisfies the Poisson equation \eqn{lapls} and gives rise to a
linear charge density $\lambda (\eta) = \eta$.
In fact, the first term produces the charge density
and the second is the first harmonic in an expansion
of the general solution for $V$ in monomials of $\r \eta$. Without either of these terms
the supergravity solution constructed from \eqn{11metric} wouldn't be possible
to even be defined. In addition, one easily checks that \eqn{vvv} is the unique potential with the
property that, besides satisfying \eqn{lapls}, it makes the overall
prefactor in \eqn{10metric} equal to unity
and hence the type-IIA supergravity solution contains the $AdS_5$ factor with
 no warping.\footnote{A solution for $V$ equal to the cubic term in $\eta$ and $\r$ appeared also in
\cite{Lin:2004nb} and gives rise to a pp-wave background. However, it corresponds to a reduction of the
continual Toda equation to the Laplace equation corresponding to a translational and not to a
rotational isometry which is our case.}

\no
Despite being of the general form  \eqn{11metric} it does not seem appropriate to directly
identify our geometry as being dual one of the ${\cal N}= 2$ quiver gauge theories
 of \cite{Gaiotto:2009we}.  One reason for this is that our geometry contains a  singularity
for $\th =\pi/2$ (or $\r=1$).\footnote{The reason for this singularity can be traced back to
the fact that the $S^3$ inside the $S^5$ which we dualised actually shrinks at $\th = \pi/2$.
This is similar to the appearance of a singularity when one performs
an abelian T-duality in $\mathbb{R}^2$ about its polar angular direction.}
Instead, we should think of the geometry we have found as capturing just part
of the general solution to \eqn{11metric}.

\no
To see how this can arise more explicitly, let us consider as a prototypical
example the Maldacena--Nunez (MN) solution \cite{Maldacena:2000mw} described
by the following potential \cite{ReidEdwards:2010qs}
\ba
&2 V_{MN}=  \sqrt{\r^{2} + (N + \eta)^{2} }- (N+\eta) \sinh^{-1}
\left( \frac{ N + \eta}{\rho} \right) \nonumber \\
&\qquad \qquad - \sqrt{\r^{2} + (N - \eta)^{2} }+ (N-\eta) \sinh^{-1}
\left( \frac{ N - \eta}{\rho} \right) \ ,
\ea
which is reproduced by the standard electrostatic solution in free three-dimensional
space with the line
density along the $\eta$-axis given by
\be
\l_{MN} = %\ha\left( |\eta+N|-|\eta-N|\right)=
\left\{
\begin{array}{cc}
                                               -N \ ,& \eta\leqslant -N \ ,\\
                                               \eta\ , &|\eta|\leqslant N \ ,\\
                                               N\ , & \eta\geqslant N\ .
                                             \end{array}
\right\}
\ee
%Let us redefine $\eta \rightarrow \frac{\eta}{N} $  and $\rho \rightarrow \frac{\rho}{N}$
%and then take a limit $N\rightarrow \infty$ whilst keeping all other coordinates fixed.
On physical grounds we expect that if we concentrate on the part of space near the line distribution
that will be equivalent to having a linear charge distribution everywhere.
Indeed, if we expand the potential for small $\eta$ and $\rho$ we find that the leading order behaviour
reproduces that in \eqn{ourpot}. In fact, this behaviour
is somewhat universal; by performing a further rescaling of $\eta$ and $\rho$
one can always tune the relative coefficient between the two harmonics to become precisely
that appearing in \eqn{ourpot} without altering the background geometry.
This is possible since nothing in the geometry depends on $V^{\prime}$ and
a linear term in $\eta$ (with constant coefficient) can be always be added at no cost.

\no
This sort of  ``zooming in'' that transforms the MN geometry to the one we
have found seems to be quite characteristic of non-abelian T-duality. Recall that,
the non-abelian T-dual of a $G_{k}$ WZW model with respect to
a subgroup $H$ gives rise to a gravitational background corresponding to
the gauged WZW model for the coset $(G_{k} \times H_{\ell})/H_{k+\ell}$ in the
limit $\ell\rightarrow \infty$  \cite{Sfetsos:1994vz} which precisely
corresponds to a zooming in part of the manifold. At the level of states
this limiting procedure can be understood as a large spin limit \cite{Polychronakos:2010nk}.
In addition, this is also reminiscent of the Penrose limit that transforms the $AdS_{5}\times S^{5}$
solution to the plane wave \cite{Blau:2002dy} and its corresponding understanding in terms of states of
parametrically large angular momentum in the gauge theory side \cite{Berenstein:2002jq}.
Due to these analogies it is natural to suggest that the geometry
we have found corresponds to a high spin sector of some parent $\cN=2$ superconformal theory.

%\no
%Note the coordinate change
%\be
%y = \eta (1-\r^2)\ ,\qq r =\r e^{\eta^2 -\r^2/2}\ ,
%\ee
%so that
%\be
%e^D = e^{\r^2- 2 \eta^2}\ .
%\ee
%The above coordinate change cannot be inverted, but we immediately see that the
%only way for $e^D$ to vanish is that $\eta\to \infty$.

\section{Concluding remarks}

In this paper we have shown how to implement non-abelian T-duality in supergravity backgrounds
supported by RR-fluxes which has been an open issue for a long time. We worked out in detail
two specific examples
by taking advantage of an $SU(2)$ subgroup of their full isometry group.
The first example was the type-IIB supergravity background
arising in the near horizon limit of the D1-D5 brane system corresponding to the $AdS_3\times S^3\times
T^4$ geometry.
We found that its non-abelian T-dual is of the form
$AdS_3 \times X_3 \times T^4$ and is a regular solution of massive IIA supergravity.
Solutions of massive IIA supergravity exist in the literature
\cite{Romans:1985tz} and more recently in \cite{Bovy:2005qq}.
However, to our best knowledge the solution we presented here is novel.
Similarly, in the second example we computed the non-abelian T-dual corresponding to
the near horizon limit of D3-branes with $AdS_5\times S^5$ geometry and found a dual
solution of the form $AdS_5 \times X_5$ in type-IIA supergravity. 
In both examples we have focused to the near horizon geometry of the corresponding brane
systems. It will be interesting to extend our construction to the full solution
expecting to obtain one-eighth and one-quarter supersymmetric solutions for the two cases,
respectively.  Equally it would be desirable to understand whether the dual geometries we find can also be understood as near-horizon limits of other brane systems.

\no
A rather generic feature of non-abelian T-duality is the appearance of non-compact variables
in the T-dual background even if the isometry group we dualise with is compact.
For the case of NS backgrounds
it was shown in \cite{Polychronakos:2010nk} that this is related to the fact that the non-abelian T-dual
effectively describes high spin sectors of some parent theory that involves as an essential ingredient the gauged WZW model action. One might hope to attempt a similar interpretation when RR-fluxes are turned on.   An intriguing feature that suggests some similar description may be possible is displayed by the non-abelian dual of the $AdS_5 \times S^5$ background.   We demonstrated that it has an eleven-dimensional origin which falls into the general class of $\cN=2$ superconformal backgrounds.  We   argued that it must effectively
describe, within the AdS/CFT correspondence, a large spin sector of the gauge theory.  Needless to say, understanding this more precisely, as well as other related
issues, will be important.

\no
Given that this is the first work in which non-abelian T-duality is
implemented in Ramond backgrounds, it would be
very interesting to provide additional examples, especially ones in which the isometry group
action is different than the special kind we focused on in this work.

\vskip .8 cm

\centerline{\bf Acknowledgments}

\no
We would like to thank P. Koerber and R. Benichou and N. Karaiskos
for helpful discussions on this research. K.S.
thanks the Galileo Galilei Institute for Theoretical Physics for
hospitality and the INFN for partial support during part of this work.
D.C.T. is by the Belgian Federal Science Policy Office through the Interuniversity Attraction
Pole IAP VI/11 and by FWO-Vlaanderen through project G011410N.

\newpage

 \appendix
\section{ Brief review of massive IIA supergravity}

In the conventions of \cite{Bergshoeff:2001pv}, the action of the massive type-IIA supergravity \cite{Romans:1985tz} is given by
\ba
S_{\rm Massive\ IIA} & = & {1\ov 2\kappa^2}\int_{M_{10}} \Bigg[e^{-2\Phi}
\left(R + 4 (\del\Phi)^2 -{H^2\ov 12}\right)  -\ha\left(m^2 + {F_2^2\ov 2} +{F_4^2\ov 4!}\right)
\nonumber\\
&&\ - \ha\left( dC_3 \wedge dC_3 \wedge B +  {m\ov 3} dC_3 \wedge B^3 + {m^2 \ov 20} B^5\right)
\Bigg]\ ,
\label{act2}
\ea
where the field strengths are defined as
\be
H= dB\ ,\qq  F_2 = dC_1 + m B \ ,\qq F_4 = dC_3 - H\wedge C_1 + {m\ov 2}B\wedge B\ ,
\ee
and where $m$ is the mass parameter. Note that, the presence of the $\ha m^2$ term in the action
reveals that $m$ plays the r\^ole of a zero-form $F_0$.
The relative coefficients have been fixed so that the field strengths are invariant
under the gauge transformations
\be
\d B= d\L \ ,\qq \d C_1 = -m \L \ ,\qq \d C_3 = - m \L \wedge B\ ,
\label{gautr}
\ee
where $\L$ is a one-form.
The Bianchi identities are
\be
dH= 0 \ ,\qq dF_2= m H \ ,\qq dF_4 = H\wedge F_2\ .
\label{biiac}
\ee
The topological term in the action can be written as
\be
-\ha \int_{M_{10}}
dC_3 \wedge dC_3 \wedge B + {m\ov 3} dC_3 \wedge B^3 + {m^2 \ov 20} B^5 = -\ha \int_{M_{11}} F_4 \wedge F_4 \wedge H \ ,
\ee
where $\del M_{11}=M_{10}$, so that gauge invariance under \eqn{gautr} becomes manifest.

\no
The equations of motions that follow from varying the metric are
\ba
&& R_{\m\n}+2 D_\m D_\n\Phi -
{1\ov 4} H^2_{\m\n}
%=e^{2\Phi}\left[
%\ha \left((F_2^2)_{\m\n} -{1\ov 4} g_{\m\n}F_2^2\right)
%+ {1\ov 12 } \left((F_4^2)_{\m\n} -{1\ov 8} g_{\m\n}F_4^2\right)
%- {m^2\ov 4}\right]\ ,
=e^{2\Phi}\Bigg[\ha (F_2^2)_{\m\n} + {1\ov 12} (F_4^2)_{\m\n}
\nonumber\\
&& \phantom{xxxxxxxxxxxxxxxxxxxxxxxxx} - {1\ov 4} g_{\m\n}
\left(\ha F_2^2 +{1\ov 24} F_4^2 + m^2\right)\Bigg]\ ,
\label{Einsteq}
\ea
whereas the dilaton equation is
\be
R + 4 D^2 \Phi - 4 (\del \Phi)^2 - {1\ov 12} H^2 = 0 \ .
\label{dilaeq}
\ee
From varying the fluxes we obtain
(after simplifying using Bianchi identities)
\ba
&& d\left(e^{-2\Phi}\star H\right) - F_2\wedge \star F_4 - \ha F_4\wedge F_4 = m \star F_2\ ,
\nonumber\\
&& d \star F_2 + H\wedge \star F_4 =0 \ ,
\label{fluxx}
\\
&&
 d \star F_4 + H\wedge  F_4 =0\ .
\nonumber
\ea
This set of equations is consistent with the Bianchi identities as it can be seen by applying
to each one of them the exterior derivative. In particular, we note the necessity of
the term proportional to $m$ in the right hand side of the first of \eqn{fluxx}.
 
 %%%%%%%%%%%%%%%%%%%%%%%%%%%%%%%%%%%%%%%%%%%%%%%%%%%%%%%%%%%%%%%%%%%%%%%%

%\bibliographystyle{JHEP}
%\bibliography{DansBib}

\providecommand{\href}[2]{#2}\begingroup\raggedright\endgroup

\end{document}